\documentclass[12pt,a4paper]{article}
\usepackage{hyperref}
\usepackage{amsmath,amsfonts,amssymb}
\usepackage[english]{babel}

\let\vphi\varphi

\let\p\partial

\let\ds\displaystyle

\begin{document}
\title{On non-Abelian Toda $A_2^{(1)}$ model and related hierarchies} %
\author{Dmitry K. Demskoi, 
Jyh-Hao Lee \\
\small{Institute of Mathematics, Academia Sinica, Taipei, Taiwan}}

\date{}
\maketitle
\begin{abstract}
We study limiting cases of the two known integrable chiral-type models with tree-dimensional configuration space. One of the initial models is the non-Abelian Toda $A_2^{(1)}$ model and the other was found by means of the symmetry approach by A.G. Meshkov and one of the authors.  The C-integrability of the reduced models is established by constructing their complete sets of integrals and general solutions. 
A description of the generalized symmetry algebras of these models is given in terms of operators mapping integrals into symmetries.
The integrals of the Liouville-type systems are known to define Miura-type transformations for their generalized symmetries. This fact allowed us to find a few new systems of the Yajima-Oikawa type. We present a recursion operator for one them.
\end{abstract}
\section*{I. Introduction}
It is known that many S-integrable hyperbolic equations have limiting cases which are integrable explicitly (see e.g. \cite{zhibsok}).
The latter have a few other characteristic properties such as presence of generalized symmetries and nontrivial integrals (pseudoconstants). These equations constitute a subclass of C-integrable equations and called the Liouville-type equations. It appears that the above mentioned interrelation between S- and C-integrable equations \cite{calog} can be used for establishing links between different though related hierarchies and also for studying their properties \cite{demliouv,demsknnl}. 

Below we consider integrable models with Lagrangians of the form
\begin{equation}
L=u_x u_t+\eta v_x w_t+f,
	\label{na_toda_lagr}
\end{equation}
where subscripts denote partial derivatives, $\eta$ and $f$ are functions of the field variables $u, v,$ and $w$. Such models are sometimes referred as chiral-type models \cite{dmm, baland}. The general form of the chiral-type Lagrangian is
\begin{equation}
	L=g_{ij}(u)u_x^i u_t^j +f(u),
	\label{lagrgen}
\end{equation}
where $g$ is some non-degenerate matrix (metric tensor). Throughout the article we assume summation over the repeated indices.

The non-Abelian Toda $A_2^{(1)}$ model has the Lagrangian
\begin{equation}
	 L_1=u_t u_x+\frac{4}{3} \frac{v_x w_t}{vw+e^u}+a \left(v w+\tfrac{3}{4} e^{u}\right)e^u+b e^{-2 u},
	 \label{m2}
\end{equation}
where $a$ and $b$ are arbitrary constants. 
Along with (\ref{m2}) we also consider a model with
\begin{equation}
	L_2=u_t u_x+4 \frac{v_x w_t}{vw+c}+a v e^u+b w e^{-u}.
	\label{m1}
\end{equation}

The reason why these models are considered in one paper is that they both related to the Yajima-Oikawa hierarchy and also because of the obvious similarity between them. The other common feature is that they both admit generalized symmetries which are {\it polynomial} w.r.t. derivatives of field variables.
The integrable model corresponding to (\ref{m2}) was derived by means of a general algebraic construction of affine non-Abelian Toda models \cite{araty}. The simplest non-Abelian Toda $A_1^{(1)}$ model is  actually the well known Lund-Regge system (complex sine-Gordon I equation) \cite{lund}. 

The model corresponding to $L_2$ was studied in \cite{demmesh}, where it was shown to have a Lax representation and infinitely many conserved densities. Obviously this model can be viewed as one of the extensions of the classical sine-Gordon model.
The limiting cases of (\ref{m1}) with $a=0$ or $b=0$ were considered in \cite{demliouv}. It was proved that in either of these two instances the model has a complete set of integrals and hence can be integrated explicitly. The actual general solution will be given in Section IV
of this paper. We also show that these reduced models are related to the open Toda $A_2$ chain. 

One of the objectives of this article is to establish connections between hierarchies of generalized symmetries \cite{olver} generated by (\ref{m2}) and (\ref{m1}). We show that they are both related to the Yajima-Oikawa hierarchy \cite{YO} by means of Miura-type transformations generated by the integrals of the reduced systems.

The paper is organized as follows. First, we consider the reduced models derived from (\ref{m2}) and (\ref{m1}), 
 construct their complete sets of integrals, proving therefore that they are of the Liouville type. The generalized Laplace invariants are used here as an auxiliary tool assisting in construction of the integrals.
 Section III 
 is devoted to finding the operators mapping integrals of systems in question into their generalized symmetries.
In Section IV
we show that solutions of the reduced systems are related to solutions of either the Liouville equation or the open Toda $A_2$ chain. This enables us to give explicit formulas of general solutions that are free of quadratures. In the last section integrals of the reduced systems are used to find integrable evolution systems related to the simplest generalized symmetries of (\ref{m2}) and (\ref{m1}).
 
%
\section*{II. Integrals and generalized Laplace invariants of reduced systems}
The Laplace invariants of systems of equations have previously been considered in a few different contexts \cite{schiefrog,zhibsok,zhibstarts,fer}. It has been established \cite{zhib} that the chains of the Laplace invariants for the most well-known hyperbolic systems having complete sets of integrals -- the open Toda chains, are finite. As we announced before the reduced systems considered in this articles are related to the open Toda $A_2$ chain. Therefore the Laplace invariants for these systems must have similar properties. There is, however, a difference: we will show that for systems with Lagrangians (\ref{m2}), (\ref{m1}) under condition $a=0$  $(b=0)$, the chains of the Laplace invariants $H_k$ terminate in classical sense, i.e. $\mbox{det}H_k\ne0, H_{k+1}=0$ for some $k$. The Laplace invariants $H_k$ for the open Toda chains are known to become degenerate though not identical zeros for some $k$ (see e.g. \cite{zhibsok}).

The Laplace invariants for the system of hyperbolic equations
\begin{equation}
u_{tx}^i=F^i(u,u_t,u_x),\ \ u=(u^1,\dots,u^n)
	\label{syshyp}
\end{equation}
are introduced as follows. 
 First, we consider the linearized system 
\begin{equation}
 S^{i}_{tx}-\frac{\p F^i}{\p u_x^j}S^j_x-\frac{\p F^i}{\p u_t^j}S^j_t-\frac{\p F^i}{\p u^j}S^j=0.
	\label{linsys}
\end{equation}
The first Laplace invariants $H_{-1},H_0$ of (\ref{linsys}) are defined as (see \cite{zhibsok} for detailed exposition)
$$
(H_0)^i_j=\frac{\p F^i}{\p u_t^k} \frac{\p F^k}{\p u_x^j}+\frac{\p F^i}{\p u^j}-D_x\left(\frac{\p F^i}{\p u_x^j}\right) ,\ \
(H_{-1})^i_j=\frac{\p F^i}{\p u_x^k}\frac{\p F^k}{\p u_t^j}+\frac{\p F^i}{\p u^j}-D_t\left(\frac{\p F^i}{\p u_t^j}\right),
$$
where the total derivative operators $D_x, D_t$ w.r.t corresponding variables are calculated in virtue of system (\ref{syshyp}).
The chain of the Laplace invariants is introduced according to the following recurrent formulas 
\begin{equation*}\begin{array}{l}
  \ds	A_{k+1}H_k=-D_t(H_k)+H_kA_k, \ \ (A_0)^i_j=-\frac{\p F^i}{\p u_x^j}, \\
	\ds (H_{k+1})^i_j=(H_k)^i_j+D_x(A_{k+1})^i_j-\frac{\p F^i}{\p u_t^s}(A_{k+1})^s_j+(A_{k+1})^i_s\frac{\p F^s}{\p u_t^j}+D_t \frac{\p F^i}{\p u_t^j}.
	\end{array}
	\label{ltransf}
\end{equation*}
If $\mbox{det}\,H_k\ne 0$, then the matrix $A_{k+1}$ and hence the next Laplace invariant $H_{k+1}$ are determined uniquely. If it is true for all $k$, then we have an infinite chain of the Laplace invariants. However, if the system in question admits nontrivial integrals, then $\mbox{det}\,H_k=0$ for some $k$. Nevertheless even in this case the chain can be continued if certain conditions are satisfied \cite{zhibsok}. 

The first Laplace invariants $H_{-1}, H_0$ for systems with Lagrangian $(\ref{lagrgen})$ under condition $f=0$ have a nice geometric interpretation, namely they are closely related to the Riemann curvature tensor. In fact the Riemann curvature tensor can be defined as the first Laplace invariant for the system
\begin{equation}
	u^i_{tx}+\Gamma^i_{jk}u_x^j u_t^k=0,
	\label{chiral}
\end{equation}
where
$$
\Gamma^i_{jk}=\frac{1}{2}g^{(is)}\left(\frac{\p g_{js}}{\p u^k}+\frac{\p g_{sk}}{\p u^j}-\frac{\p g_{jk}}{\p u^s}\right), \ \ g^{(is)}=\frac{1}{2}(g^{is}+g^{si})
$$
Note that $\Gamma^i_{jk}$ is not necessarily symmetric. It is not difficult to find that for system (\ref{chiral}) the Laplace invariant $H_0$ is given explicitly by
$$
(H_0)^i_j=\left(\frac{\p \Gamma^i_{js}}{\p u^k}-\frac{\p \Gamma^i_{ks}}{\p u^j}+\Gamma^i_{kp}\Gamma^p_{js}-\Gamma^i_{jp}\Gamma^p_{ks}\right)u_x^k u_t^s=R^i_{skj}u_x^k u_t^s,
$$
where $R^i_{skj}$ is the curvature tensor corresponding to $g_{ij}$.
It immediately follows from this that $\mbox{det} H_0=0$. Indeed, due to the antisymmetry of $R^i_{skj}$ w.r.t. indices $k,j$ we have $R^i_{skj}u_t^s u_x^k u_x^j=0$. This is a reflection of the well known fact that any system (\ref{chiral}) has the first order integral
\begin{equation}
	\omega=~g_{ij}u_x^iu_x^j,\ \ D_t\omega=0.
	\label{firstint}
\end{equation}
{ Generally we conjecture that if $n>k_0$, then a coupled system of form (\ref{syshyp}) admits $n-k_0$ first order integrals. Furthermore if $\mbox{rank}\, H_i=k_i$ and $k_{i-1}>k_i$ for $i>0$, then system (\ref{syshyp}) admits $k_{i-1}-k_i$ integrals of the order $i+1$. This statement has been verified for different Liouville-type systems and in particular for the open Toda chains \cite{zhib}. We have also verified the validity of this statement for the reduced systems in question. }

The explicit form of the system corresponding to Lagrangian (\ref{m2}) is
\begin{equation}
\begin{array}{l}
	u_{tx}=-\frac{2}{3} \psi^2 v_x w_t e^u+\frac{1}{2}\,a\, e^u(\frac{3}{2} e^{u}+ v w)-b\, e^{-2 u},\\[2mm]
	v_{tx}=\psi v_x (w v_t+e^u u_t)+\frac{3}{4}\, a \psi^{-1} v e^u, \ \
	w_{tx}=\psi w_t (w_x v+e^u u_x)+\frac{3}{4}\, a \psi^{-1} w e^u,
\end{array}\label{asys}
\end{equation}
where  $\psi=(vw+e^u)^{-1}.$ 
Below we consider the reduced systems derived from (\ref{asys}) by successively setting $a=b=0$, $a=0$, and $b=0$. The corresponding hyperbolic systems will be referred as ${\cal S}_1, {\cal S}_1^b$, and ${\cal S}_1^a$. Systems  ${\cal S}_1^a$ and ${\cal S}_1^b$ are also known as the reduced $A_2^{(2)}$ Bershadsky-Polyakov and $A^{(1, 1)}_2$ non-Abelian Toda models (see \cite{sotk} and references therein).
Our objective is to show that these systems have terminating sequences of the Laplace invariants and complete sets of integrals. 

The system corresponding to (\ref{m1}) has the form
\begin{equation}\begin{array}{l}
u_{tx} = \frac{a}{2} v e^u-\frac{b}{2} w e^{-u},\ \
v_{tx} = \frac{b}{4}\varphi^{-1} e^{-u}+\varphi w v_x v_t, 
w_{tx} = \frac{a}{4}\varphi^{-1} e^{-u}+\varphi v w_t w_x,
\end{array}
	\label{asysm1}
\end{equation}
where $\varphi=(vw+c)^{-1}$.
The reduced systems derived from (\ref{asysm1}) were considered in \cite{demliouv} where they were shown to belong to a class of Liouville-type systems. By analogy with the previous systems they will be referred as ${\cal S}_2, {\cal S}_2^b$, and ${\cal S}_2^a$. { Note that systems (\ref{asys}) and (\ref{asysm1}) admit the following symmetries $t\to x,x\to t, v \to w, w \to v$ and $t\to x, x\to t$ correspondingly. This allows one to construct $t-$integrals from $x-$integrals and vice versa if either of them is known.}
 
We start with the fully reduced system ${\cal S}_1$ $(a=b=0)$
\begin{equation}\begin{array}{l}
u_{tx} = -\frac{2}{3}\psi^2 e^u v_x w_t, \ \	
v_{tx} = \psi v_x (e^{u} u_t+v_t w), \ \
w_{tx} = \psi w_t (e^{u} u_x+w_x v).
\end{array}
	\label{a12deg}
\end{equation}
It is not difficult to check that for this system we have $\mbox{rank}\,H_0=1$ and $H_1 H_0=0$, therefore we conjecture that (\ref{a12deg}) has two first order and one second order integrals. To derive the complete set of integrals for this system, we used the procedure suggested in \cite{demliouv}. For the applicability of the procedure one needs to have a nontrivial integral and a non-degenerate higher commuting flow (symmetry). The nontrivial integral for (\ref{a12deg}) has the form (\ref{firstint}). The simplest generalized symmetry is common for all systems derived from (\ref{m2}) and given by formula (\ref{comsymm}) (see below).

It is convenient to write integrals in terms of the following quantities
\begin{equation}
	\alpha=u, \ \ \beta=v_x w\psi, \ \ \gamma=\ln w.
	\label{trans0}
\end{equation}
Then the integrals of (\ref{a12deg}) can be written as 
\begin{equation}
	\begin{array}{l}
	 m=\alpha_x+\frac{2}{3}\beta,\ \  
	  p=\frac{4}{3}\beta(\alpha_x- \gamma_x+\frac{1}{3}\beta), \ \  
	 q=2(\alpha_x+\gamma_x-\frac{1}{3}\beta -\beta^{-1}\beta_x). 
	\end{array}
	\label{ints}
\end{equation}
Note that integral (\ref{firstint}) can be brought into the form ${\cal \omega}=m^2-p$.

System ${\cal S}_1^b$ $(a=0)$:
\begin{equation}\begin{array}{l}
u_{tx} = -\frac{2}{3}\psi^2 e^u v_x w_t-b e^{-2 u}, \ \	
v_{tx} = \psi v_x (e^{u} u_t+v_t w), \ \
w_{tx} = \psi w_t (e^{u} u_x+w_x v).
\end{array}
	\label{a13deg}
\end{equation}
For this system we have $\mbox{rank}\,H_0=3$ and $H_1=0$ so we look for the three independent second order integrals.
Because system ${\cal S}_1$ appears to be a limiting case of ${\cal S}_1^a$ and ${\cal S}_1^b$ the integrals of the latter can be expressed in terms of integrals of the former.
By direct calculation it is not difficult to find that the integrals are
\begin{equation}
	\begin{array}{l}
		\mu=m-\frac{1}{2} q,\ \ \nu=m_x+m^2-p,\ \	\lambda=-2 p_x-q p.	
	\end{array}
	\label{S2int}
\end{equation}

System ${\cal S}_1^a$ $(b=0)$:
\begin{equation}
\begin{array}{l}
u_{tx}=-\frac{2}{3}\psi^2 e^{u} v_x w_t+\frac{a}{4} e^u(e^u+2\psi^{-1}), \\[1mm]
v_{tx}=\psi v_x (e^u u_t+w v_t)+\frac{3}4 a v e^u \psi^{-1}, \ \
w_{tx}=\psi w_t (e^u u_x+w_x v)+\frac{3}{4} a w e^u \psi^{-1}.
\end{array}
\label{a14deg}
\end{equation}
As in the previous case we have $\mbox{rank}\,H_0=3$, $H_1=0$ and thus the system has three second order integrals given explicitly by
\begin{equation}
	\begin{array}{l}
	\rho=m^2-m_x-p,\ \
	\theta=p\, q, \ \
	\phi=p_x p^{-1}+\frac{1}{2} q-m.
	\end{array}
	\label{rtphi}
\end{equation}
The complete set of integrals for ${\cal S}^a_2$ and hence for ${\cal S}^b_2$ were constructed in \cite{demliouv}. 
As in the previous case it is convenient to introduce the quantities
\begin{equation}
	\alpha=u,\ \ \beta=v_x w\vphi, \ \ \gamma=\ln w.
	\label{abg}
\end{equation}
System ${\cal S}_2$ decouples into the d'Alambert equation and the reduced Lund-Regge system
\begin{equation}
	u_{tx}=0, \ \ v_{tx}=w\vphi v_t v_x, \ \ w_{tx}=v\vphi w_t w_x. 
	\label{LRred}
\end{equation}
The integrals are
\begin{equation}
	m=-\tfrac{1}{2}\,\alpha_x, \ \ p=-\beta \gamma_x,\ \ q=2(\gamma_x-\alpha_x-\beta_x \beta^{-1}-\beta).
	\label{intermed2}
\end{equation}
System (\ref{LRred}) was used in \cite{sokstarts,demstarts} as the working example of a Liouville-type system.

System ${\cal S}_2^a$ is given by
\begin{equation}
	u_{tx} = \tfrac{a}{2} v e^u, \ \ v_{tx} = \varphi w v_x v_t, \ \ w_{tx} = \tfrac{a}{4}\psi^{-1} e^u+\varphi v w_t w_x.
	\label{sysa}
\end{equation}
The complete set of integrals for this system has the form (\ref{S2int}) with $m,p,$ and $q$ given by (\ref{intermed2}).
System ${\cal S}_2^b$ is obtained from (\ref{sysa}) by means of the transformation 
$u\to~-u,$ $v\to~w,$ $w\to~v,$ $a\to~b.$
The integrals for this system have the form (\ref{rtphi}) with $m,p,$ and $q$ given by (\ref{intermed2}). We would like to point out that the presented sets of integrals are {\it minimal}.
\section*{III. The structure of generalized symmetries} \label{structure}
Higher symmetries of the Liouville-type systems are known to have the special structure
$$
S=M \omega,
$$
where $M$ is some linear differential operator and $\omega$ a vector-function of integrals. Function $S$ is assumed to satisfy equation (\ref{linsys}).
Operator $M$ gives a complete description of symmetry algebra for a given Liouville-type hyperbolic system. It satisfies the following operator equation 
\begin{equation}
	(D_x D_t-F_*)M= T D_t,
	\label{mrel}
\end{equation}
where $T$ is some differential operator and $F_*$ stands for the Freshet derivative of the right hand side of (\ref{syshyp}). In principle, relation (\ref{mrel}) can be used to find operator $M$, but it appears more convenient to use results of \cite{demstarts} where it was proved that for any Liouville-type system of the form (\ref{syshyp}) there exists a differential operator $P$ such that
\begin{equation}
	\omega^+_*=(-D_x+F_{u_t})^+\circ P,
	\label{mform}
\end{equation}
where $\omega^+_*$ stands for the operator formally conjugated to $\omega_*$. Then according to \cite{starts} the operator $M=g_s^{-1}P$ maps integrals of (\ref{syshyp}) into its symmetries, where $g_s$ is the symmetric part of the metric tensor. The matrix $g_s^{-1}$ for models ${\cal S}_1$, ${\cal S}^a_1$, and ${\cal S}^b_1$ has the form
$$g_s^{-1}=\left(
\begin{array}{ccc}
1 & 0 & 0 \\
0 & 0 & \frac{3}{2}\psi^{-1} \\
0 & \frac{3}{2}\psi^{-1} & 0
\end{array}\right).
$$
Having found integrals for these models, it is not difficult to factorize operator $\omega^+_*$ as in formula (\ref{mform}), and thus to find the operator $P$. Therefore we have found the following $M-$operator for model (\ref{a12deg})
$$
{\cal M}=g_s^{-1}P=
\left(\begin{array}{ccc}
	1 & -\frac{2}{3}\psi w v_x & 2 \\
	0 & v_x & 3 v \\
	w & w_x-w u_x-\frac{2}{3}w^2 v_x\psi& 3 (\psi v_x)^{-1}D_x-w
\end{array}\right).
$$
We denote ${\cal M}_a$ and ${\cal M}_b$ the M-operators for models ${\cal S}_1^a$ and ${\cal S}_1^b$ correspondingly. These operators can be factorized in the following way
\begin{equation}
	{\cal M}_a={\cal M} {\cal F}_a, \ \ {\cal M}_b={\cal M} {\cal F}_b
	\label{mfactor}
\end{equation}
where
$$
{\cal F}_b=\left(\begin{array}{ccc}
-2&D_x-2 m&0 \\
0&-2&D_x-\frac{1}{2}q\\
1&0&\frac{1}{4}p
\end{array}\right),\ \
{\cal F}_a=\left(
\begin{array}{ccc}
D_x+2 m&0&1 \\
2&2 q&-2 p^{-1}D_x\\
0&-p&-\frac{1}{2}
\end{array}\right).
$$
If we denote as ${\cal A}_1$, ${\cal A}^a_1$, and  ${\cal A}^b_1$ the generalized symmetries algebras of systems ${\cal S}_1$, ${\cal S}^a_1$, and  ${\cal S}^b_1$  correspondingly, then from (\ref{mfactor}) the following relation follows
$
{\cal A}^a_1 \subset {\cal A}_1, \ \ {\cal A}^b_1 \subset {\cal A}_1 .
$
Finally we would like to point out the the simplest generalized symmetry of (\ref{asys}) can be written in the form
\begin{equation}
	\left(\begin{array}{c}
	u_t\\ v_t\\ w_t
	\end{array}
	\right)={\cal M}_a\left(\begin{array}{c}
	0\\ 1\\ 0
	\end{array}
	\right)=
	{\cal M}_b\left(\begin{array}{c}
	0\\ 0\\ -4
	\end{array}
	\right).
	\label{comsymm}
\end{equation}
It is easy to check that (\ref{comsymm}) is the second order polynomial (w.r.t derivatives) evolutionary system.
\section*{IV. General solutions of the reduced systems} \label{solut}
In \cite{leznov} a reduction procedure was used to find a general solution of the open Toda $A_n$ chain. The idea is to replace a system in question by an equivalent higher order PDE which can then be integrated explicitly. Here a similar procedure is applied to systems ${\cal S}^a_i$, ${\cal S}^b_i$, and ${\cal S}_i$ ($i=1,2$). The solutions of these systems will be given in quadrature-free form.

We start with the simplest system ${\cal S}_1$ which is given by (\ref{a12deg}). 
First, expressing $u_t$ and $u_x$ from the second and third equations correspondingly and then
calculating the compatibility condition $u_{tx}=u_{xt}$ we find
$$
h_{tx} h-h_t h_x=0,
$$
where $h=w_t v_x^{-1}$.
This allows us to parametrize functions $v$ and $w$ the following way
$$ v=T_1 e^{s} s_t, \ \ w=-X_1 e^{s} s_x.$$
Here and below $T_i(t)$ and $X_i(x)$ are arbitrary functions of the indicated variables.
Now considering equations $(\ref{a12deg})_2,(\ref{a12deg})_3$ as ODEs (w.r.t. to $u$) we find
$$ u = \ln(T_1 X_1)+2 s+\ln(-s_{tx}). $$
Substituting this expression into $(\ref{a12deg})_1$ we find that $s$ satisfies the equation
\begin{equation}
	s_{txx}s_{ttx}-s_{ttxx} s_{tx}=\tfrac{8}{3}\,s_{tx}^3.
	\label{wr1}
\end{equation}
Again the substitution $s_{tx}=-\exp(r)$
reduces (\ref{wr1}) to the Liouville equation
$$
r_{tx}=\tfrac{8}{3}\, e^r
$$
having the well-known general solution
$$
r=\ln\left(\tfrac{3}{4}\tfrac{X_2'T_2'}{(X_2+T_2)^2}\right).
$$
Thus we have finally
$$
s=\tfrac{3}{4}\ln(X_2+T_2)+T_3+X_3.
$$
Therefore the general solution of (\ref{a12deg}) is
$$
\begin{array}{l}
u= 2 (T_3+ X_3)+\ln\left(\frac{3}{4}\frac{T_1 X_1 X_2'T_2'}{\sqrt{X_2+T_2}}\right), \ \
v= T_1 \exp(X_3+T_3)(X_2+T_2)^\frac{3}{4}\left(\frac{3}{4}\frac{T_2'}{X_2+T_2}+T_3'\right), \\[2mm]
w= -X_1 \exp(X_3+T_3)(X_2+T_2)^\frac{3}{4}\left(\frac{3}{4}\frac{X_2'}{X_2+T_2}+X_3'\right).
\end{array}
$$
The same reduction procedure can be applied to system ${\cal S}^b_1$. The difference is that instead of (\ref{wr1}) one gets the following system
$$
s_{txx}s_{ttx}-s_{ttxx} s_{tx}-\tfrac{8}{3}\,s_{tx}^3=b \exp(r-4 s), \ \ r_{tx}=0
$$
which in turn is equivalent to 
$$
s_{tx}=\exp(-\tfrac{8}{3} s+ \tau),\ \ \tau_{tx} = -b\exp(\tfrac{4}{3} s-2 \tau+r), \ \ r_{tx}=0.
$$
It is easy to see that the latter system is reducible to the open $A_2$ Toda chain by the transformation $s\to 3/4 s-3/20 r,\, \tau\to \tau-2/5 r$.
Using this connection \cite{leznov} one can express the general solution of ${\cal S}_1^b$ in the form
$$
u=\ln(r_t r_x)+\log(-s_{tx})+2 s, \ \ v=r_t \exp(s) s_t, \ \ w=-r_x \exp(s) s_x,
$$
where
\begin{equation}
r = \ln\left(s_{txx} s_{ttx}-s_{ttxx} s_{tx}-\tfrac{8}{3} s_{tx}^3\right)+4 s-\log(b), \ \
s=\tfrac{3}{4}\ln(X_1 T_1+X_2 T_2+X_3 T_3).	
\end{equation}

The general solution of the model ${\cal S}^a_1$ (with $a=-4/3$) can be obtained from the solution of ${\cal S}^b_1$ by using the transformation \cite{sotk}
$$
u \to -u-\tfrac{1}{2}\ln\left(1+\tfrac{4}{3}e^{-u} v w\right),\ \ v \to w e^{-u}(1+\tfrac{4}{3} e^{-u}v w)^{-1/4},\ \ 
w = \tfrac{4}{3} v e^{-u}(1+\tfrac{4}{3} e^{-u} v w)^{-1/4} . 
$$

Now consider the reduced systems ${\cal S}_2$ and ${\cal S}_2^a$, i.e. (\ref{LRred}) and (\ref{sysa}). The general solution of ${\cal S}_2$ can be found the following way. First, we express $w$ from the first equation in (\ref{LRred})
$$
w=v_{tx}(v_{tx}v-v_t v_x)^{-1}
$$
and then substituting it into the second equation, we get
\begin{equation}
	\mbox{det}\left(\begin{array}{ccc}
	v & v_t & v_{tt}\\
	v_x & v_{tx} & v_{ttx} \\
	v_{xx} & v_{txx} & v_{ttxx}
	\end{array}\right)=0.
\end{equation}
The latter equation has the following general solution $v=X_1 T_1+X_2T_2$, and thus we have $$w=\frac{(X_1'T_1'+X_2'T_2')}{(T_2T_1'-T_1T_2')(X_1X_2'-X_2X_1')}.$$

Now we turn to system ${\cal S}_2^a$. The variables $v$ and $w$ can be expressed from the first and second equations in (\ref{sysa}), this gives
\begin{equation}
	v = 2 a^{-1} \exp\left(-\tfrac{s}{2} \right),\ \   w = \tfrac{a c}{4} \exp\left(\tfrac{s}{2}\right) (s_x s_t-2 s_{tx})s_{tx}^{-1},
	\label{vw}
\end{equation}
where 
$ s = 2 u-2 \ln u_{tx}$.
Substituting (\ref{vw}) to $(\ref{sysa})_3$ yields
\begin{equation}
	s_{ttxx} s_{tx}-s_{ttx} s_{txx}-s_{tx}^3=-\tfrac{1}{2} s_{tx}^2\exp\left(u-\tfrac{1}{2} s\right).
\end{equation}
If we introduce the new quantity $r = -2 \ln(s_{tx})-u+2 s$, then one can check that $r_{tx}=0$. Therefore system (\ref{sysa}) is equivalent to the Open Toda $A_2$ chain coupled with the d'Alambert equation
\begin{equation}
	u_{tx}=\exp\left(u-\tfrac{1}{2}s\right),\ \ s_{tx}=\exp(-\tfrac{1}{2} u+s-\tfrac{1}{2} r), \ \ r_{tx}=0.
	\label{sysaa2}
\end{equation}
This enables us to express the general solution of (\ref{sysaa2}) in the form
\begin{equation}
\begin{array}{c}
	\ds u=-2\log Q, \ \  v=\tfrac{4}{a} (Q_x Q_t-Q_{tx}Q), \\[2mm] 
	\ds w=-\tfrac{a c}{4}\frac{Q_{ttxx}Q-Q_{tt}Q_{xx}}{Q (Q_{xx} Q_{tt} Q_{tx}+Q_x Q_t Q_{ttxx}-Q Q_{tx} Q_{ttxx}-Q_{txx} Q_x Q_{tt}-Q_{ttx} Q_t Q_{xx}+Q_{ttx} Q Q_{txx})} ,
	\end{array}
	\label{sysagensol}
\end{equation}
where 
$$Q=X_1 T_1+X_2 T_2 +X_3 T_3.$$
\section*{V. Differential substitutions and modified Yajima-Oikawa systems}\label{msystems}
It is well known \cite{zhibsok} that the minimal integrals of Liouville-type systems define differential substitutions for their generalized symmetries. Therefore having constructed them for ${\cal S}_i, {\cal S}^a_i$, and ${\cal S}^b_i$ we also found the differential substitutions for generalized symmetries of these systems. Now it is easy task to construct corresponding modified evolutionary systems for (\ref{comsymm}), but first let us rewrite (\ref{comsymm}) in variables (\ref{trans0}):
\begin{eqnarray}
\lefteqn{\begin{array}{ll}
	\alpha_\tau =   -\frac{2}{3}\beta_x+\frac{4}{3}\beta\alpha_x, &
	\beta_\tau =   \beta_{xx}-2(\gamma_x\beta)_x +\frac{4}{3}\beta_x\beta , \\[1mm]
	& \gamma_\tau =  -\gamma_{xx}+\alpha_{xx}+\alpha_x^2-\gamma_x^2 -\frac{1}{3}\beta^2+\frac{4}{3}\beta\gamma_x.
\end{array}	\label{modif1}}{\qquad\qquad\qquad\qquad\qquad\qquad\qquad\qquad\qquad\qquad}
\end{eqnarray}
The second system in the chain of transformed systems corresponds to (\ref{ints}) and is given by
\begin{eqnarray}
\lefteqn{\begin{array}{ll}
		m_\tau=p_x,&  p_\tau=-p_{xx}-(p q)_x+2 m p_x, \\[1mm]
	& q_\tau=q_{xx} +2 (m q)_x-\frac{1}{2}(q^2)_x.
	\end{array}
	\label{myo1}}{\qquad\qquad\qquad\qquad\qquad\qquad\qquad\qquad\qquad\qquad}
\end{eqnarray}
The integrals of ${\cal S}^a_1$ and ${\cal S}^b_1$ take system (\ref{myo1}) into
\begin{eqnarray}
\lefteqn{\begin{array}{ll}
\rho_\tau= \theta_x, & \theta_\tau=\theta_{xx} -2 (\phi\theta)_x, \\
& \phi_\tau=-\phi_{xx}-(\phi^2-\rho)_x, 
\end{array}\label{yo}}{\qquad\qquad\qquad\qquad\qquad\qquad\qquad\qquad\qquad\qquad}\\[2mm]
\lefteqn{\begin{array}{ll}
		\nu_\tau=-\lambda_x, & \lambda_\tau = -\lambda_{xx}+2(\lambda\mu)_x, \\
		& \mu_\tau=\mu_{xx}  +(\mu^2-\nu)_x
	\end{array}\label{nml}}{\qquad\qquad\qquad\qquad\qquad\qquad\qquad\qquad\qquad\qquad}
\end{eqnarray}
correspondingly.
Note that systems (\ref{yo}) and (\ref{nml}) are related by the transformation $\tau \to - \tau$. 
System (\ref{yo}) is known (see e.g. \cite{demliouv}) to be related to the Yajima-Oikawa system \cite{YO}
\begin{eqnarray}
\lefteqn{\begin{array}{ll}
U_\tau=(V W)_x, & V_\tau=-V_{xx}+U V,\\
& W_\tau=W_{xx}-U W
\end{array}}{\qquad\qquad\qquad\qquad\qquad\qquad\qquad\qquad\qquad\qquad}
	\label{yajima}
\end{eqnarray}
by the transformation
\begin{equation}
	\rho=U, \ \ \theta=V W,\ \ \phi=V_x/V.
	\label{yoyo}
\end{equation}
On the other hand, it is known \cite{demliouv} that the simplest generalized symmetry 
\begin{equation}
	\begin{array}{ll}
		u_\tau=2\vphi v_x w_x, & v_\tau=v_{xx} -2 \vphi {v v_x w_x}+ u_x v_x, \\
		& w_\tau=-w_{xx}+2 \vphi{w v_x w_x}+ u_x w_x
	\end{array}\label{sysasym}
	\end{equation}
	of (\ref{asysm1}) is related to (\ref{myo1}) by means of the integrals of ${\cal S}_2$.
Note that system (\ref{sysasym}) in variables $\alpha, \beta$, and $\gamma$ has the simple polynomial form \cite{dmm}
\begin{equation}
\begin{array}{ll}
\alpha_\tau=2\beta \gamma_x, & \beta_\tau=\beta_{xx}+(\beta^2+\beta\alpha_x-2\beta\gamma_x)_x, \\
& \gamma_\tau=-\gamma_{xx}+\alpha_x\gamma_x-\gamma_x^2+2\beta\gamma_x.
\end{array}
\label{sysamar}
\end{equation}
{The evolution systems listed above constitute a subclass of modified Yajima-Oikawa systems singled out by the relation to hyperbolic systems with Lagrangians (\ref{m2}) and (\ref{m1}). Their integrability obviously follows from the integrability of the Yajima-Oikawa system. Many of their properties like bi-Hamiltonian structure, recursion operators, etc, can be obtained from the ones of the Yajima-Oikawa system itself. The most interesting of them is probably system (\ref{myo1}) as it is related to both hierarchies of systems with (\ref{m2}) and (\ref{m1}). We found its recursion operator in the form
\begin{equation}
\begin{array}{l}
	R=
	\left(\begin{array}{ccc}
	\frac{1}{4} & 0 & 0 \\
	0 & 1 & 0 \\
	0 & 0 & 1
	\end{array}\right)D_x^2+\left(\begin{array}{ccc}
	0 & -\frac{3}{4} & 0 \\
	0 & q-2m & 0 \\
	\frac{3}{2}q & 0 & 2m-q
	\end{array}\right)D_x\\
	\qquad +\left(\begin{array}{ccc}
	p-m^2 & \frac{1}{2}m-\frac{1}{4}q & -\frac{3}{4}p \\
	-\frac{3}{2}pq-\frac{3}{2}p_x & p-mq+\frac{1}{2}q_x+\frac{1}{4}q^2 & p_x-2 m p+p q \\
	-\frac{1}{2}q^2+mq+\frac{5}{2}q_x & \frac{3}{2}q & -\frac{3}{2}q_x+2 m_x+p-m q+\frac{1}{4}q^2
	\end{array}\right)\\
	\qquad - 
	\left(
	\begin{array}{c}
	m_x\\
	p_x\\
	q_x
	\end{array}\right) D_x^{-1}
\left(
	\begin{array}{ccc}
	m & -\frac{1}{2} & 0
	\end{array}
	\right)
+\left(
	\begin{array}{c}
	p_x\\[2mm]
	-p_{xx}-p_x q-q_x p+2 p_x m\\
	q_{xx}+2 q m_x+2 q_x m-q q_x
	\end{array}\right)  D_x^{-1} \left(
	\begin{array}{ccc}
	1 & 0 & -\frac{1}{2}
	\end{array}
	\right).
	\end{array}
	\label{RM}
\end{equation}
One can see that it has standard structure of the nonlocal part, i.e. it is a product of symmetries and co-symmetries. We would also like to point out that (\ref{myo1}) is Hamiltonian with the following {\it local} Hamiltonian operator
\begin{equation}
	\left(\begin{array}{c}
	m_t\\ p_t\\ q_t
	\end{array}
	\right)=J \left(\begin{array}{c}
	\frac{\delta H_1}{\delta m}\\[1mm] \frac{\delta H_1}{\delta p}\\[1mm] \frac{\delta H_1}{\delta q}
	\end{array}
	\right),
\end{equation}
where 
\begin{equation}
J=
\left(
	\begin{array}{ccc}
	-\frac{1}{2}D_x&0&-2 D_x\\
	0&2 p D_x+p_x&2 D_x^2+(q-4 m)D_x\\
	-2 D_x&-2 D_x^2+D_x(q-4 m)&0
	\end{array}
\right),
\label{locham}
\end{equation}
}
$$
H_1=-\tfrac{1}{2}p\, q
$$
Applying (\ref{RM}) to (\ref{locham}) one can generate infinitely many Hamiltonian operators.

Finally we would like to note that in paper \cite{araty} one more system of the Yajima-Oikawa type is found. It has the form
\begin{equation}
	\begin{array}{l}
	u_t=-(w v)_x, \\
	v_t=v_{xx}-v u_x-v u^2-w v^2, \\
	w_t=-w_{xx}-w u_x+w u^2+w^2 v.
	\end{array}
	\label{sotkyaj}
\end{equation}
The Miura-type transformation relating (\ref{sotkyaj}) with (\ref{yajima}) is given by
$$
U=u^2+u_x+v w, \ \ V=w_x+w u,\ \ W=-2 v.
$$
\section*{Acknowledgements}
Authors are grateful to M.V.Pavlov, O.K.Pashaev, and V.V.Sokolov for fruitful discussions.

\end{document}